# Transport coefficients in composites

Kamen Dimitrov, Boryan Radoev and Roumen Tsekov
Department of Physical Chemistry, University of Sofia, 1164 Sofia, Bulgaria

The Maxwell approach from electrostatics is applied for calculation of transport coefficients in composites. The viscosity of a dilute emulsion is obtained as a function of the volume fraction of dispersed phase. The derived new formula is asymptotically correct and more general than the linear relationships usually used. The method is applied also for description of the influence of fluctuations on the transport coefficients.

The theory of transport coefficients in composites is arisen and developed in connection of particular problems in hydrodynamics and electrostatics. This circumstance, combined by the heavy ponderous mathematical techniques typical for such kind of problems, has impeded the formulation of a general approach and its further development. Usually, the transport in composites is considered as a transfer of a field $\mathbf{v}^0$ (hydrodynamic, electrical, thermal, etc.) in a medium with a transport coefficient $\eta^0$ in the presence of a dispersed phase with a transport coefficient $\eta$. In the simplest case the dispersed phase particles are uniform spheres with radius $R$. These particles induce a perturbation field $\mathbf{v}'$ and thus the resulting field equals to $\mathbf{v} = \mathbf{v}^0 + \mathbf{v}'$. The aim is to idealize the composite as a homogeneous phase with an effective transport coefficient $\eta_{eff}$. The replacement of the real system $(\mathbf{v}, \eta^0, \eta, R)$ by an idealized one $(\mathbf{v}^0, \eta_{eff})$ reduces the problem to determination of $\eta_{eff}(\eta^0, \eta, R)$. In the literature three approaches to this problem are developed:

(i) Expressing of the perturbation field, induced by an ensemble of spheres, by the perturbation of an effective single sphere. This idea belongs to Maxwell [1] and historically it is the first method for calculation of effective transport coefficients. Originally it has been applied to the Ohmic resistance of composites;

(ii) Juxtaposition of fluxes in the real and idealized systems. This idea is proposed by Burgers [2], who applied it for obtaining the effective viscosity of suspensions. This method has been further developed and systematically applied by Landau and Lifshitz in hydrodynamics [3] and electrostatics [4];

(iii) Comparison of energy consumption in the real and idealized systems. This is the well-known Einstein method in hydrodynamics balancing the energy of dissipation [5]. In optics one could consider the extinction as an effective property.

The Maxwell approach has some advantages in formal aspects as well as in the possibility for generalization. These features are demonstrated in the present study by obtaining a general expression for the effective viscosity of dispersions as well as in the study of the role of fluctuations on the transport coefficients.

## Viscosity of dilute heterogeneous systems

Let us consider uniform spherical droplets with radius $R$ and viscosity $\eta$ dispersed in a liquid with viscosity $\eta^0$. An external deformation velocity field $\mathbf{v}^0 = \mathbb{D}^0 \cdot \mathbf{r}$ acts on this emulsion. Since the fluid is incompressible, $\nabla \cdot \mathbf{v}^0 = 0$, the constant symmetric tensor $\mathbb{D}^0$ should possess a zero trace. Each droplet causes a Stokes field $\mathbf{v}'$ around itself decreasing asymptotically at large distances as a dipole one

$$\mathbf{v}'(r \to \infty) \to 2.5\gamma R^3 \mathbf{r} \cdot \mathbb{D}^0 \cdot \mathbf{rr}/r^5 \tag{1}$$

where $r = |\mathbf{r}|$ is the distance from the droplet center and $\gamma$ is a coefficient reflecting the interaction of the fields inside and outside of the droplet. All types of such interactions lie in the frames of two limit models for immiscible and miscible fluids, respectively, [6]

$$\gamma = (\bar{\eta} + 2.5)/(\bar{\eta} + 1) \qquad \gamma = (\bar{\eta} - 1)/(\bar{\eta} + 1.5) \tag{2}$$

where $\bar{\eta} \equiv \eta/\eta^0$. The difference between these two models is in the transition of the stress tensor across the droplet surface. In the case of immiscible fluids the stress tensor components jump due to a surface tension, while for miscible fluids the stress tensor is continuous since the surface tension is zero. The widely employed case of suspensions of solid spheres in a viscous fluid corresponds to interaction coefficient $\gamma = 1$ since $\eta \to \infty$.

Let us consider a hypothetical sphere with radius $R_{eff}$ placed in the emulsion and containing $n$ smaller droplets. This ensemble of droplets will cause a perturbation field $\mathbf{v}' = \sum \mathbf{v}'_i$. Following the Maxwell approach the same sphere is idealized as a droplet of homogeneous fluid with viscosity $\eta_{eff}$ placed in the original surrounding pure fluid. Note that the idealized sphere should be considered as miscible one as far as no physical boundary with the surrounding fluid exists. Let us mark the perturbation field of the effective sphere as $\mathbf{v}'_{eff}$. Then according to the potential theory at large distances the two perturbation fields should coincide

$$\mathbf{v}'_{eff}(r \to \infty) = \sum \mathbf{v}'_i(r \to \infty) \tag{3}$$

since neutrally buoyant particles (particles moving along the streamlines without delay) generate at low Reynolds number dipole fields at large distances [7], i.e. $\mathbf{v}'(r \to \infty) \to 1/r^2$ as expressed in Eq. (1). The next step is to consider the resultant field of a droplet in ensemble as the perturbation field of a single droplet. Generally, between the droplets exist hydrodynamic interactions and then the external field $\mathbb{D}$ acting on a single particle will be perturbed by the generated fields of the neighboring droplets. As discussed below, the Maxwell approach succeeds to account in part for these interactions by the help of the unperturbed external field, i.e. via $\mathbb{D}^0 = \mathbb{D}(\phi \ll 1)$ where $\phi = nR^3/R_{eff}^3$ is the volume fraction of the dispersed phase in the emulsion. At higher concentrations $\phi \approx 1$ there are some restrictions for applying the asymptotic Eq. (1) which will be discussed later.

For relatively diluted emulsions of uniformly distributed droplets, the effective perturbation field from Eq. (3) can be expressed by the perturbation field caused by a single droplet $\mathbf{v}'_1$, $\mathbf{v}'_{eff}(\phi \ll 1) = n\mathbf{v}'_1$. Introducing Eq. (1) in this relation yields

$$\gamma_{eff}(\overline{\eta}_{eff}) = \gamma(\overline{\eta})\phi \qquad (4)$$

with $\overline{\eta}_{eff} \equiv \eta_{eff}/\eta^0$. Note that the interaction coefficient of the single droplet $\gamma$ obeys the first expression from Eq. (2), while the coefficient $\gamma_{eff}$ of the effective sphere obeys the second one. Using them it leads to the final result for the effective viscosity of dilute emulsion

$$\overline{\eta}_{eff} = 1 + 2.5\gamma\phi/(1-\gamma\phi) \qquad (5)$$

This relation contains many of the existing in the literature results of dilute emulsions [5, 8-10]. As is seen, it is more general than the Einstein approximate formula $\overline{\eta}_{eff} = 1 + 2.5\gamma\phi$. The nature of the non-linear dependence in Eq. (5) is the same as that in the well-known Lorentz-Lorenz or Clausius-Mossotti equation discussed later [see Eq. (15)]. The collective effects can be accounted for also by consecutive 'dilution' of the droplets in an effective environment. Since the volume fraction of a single droplet $\phi_1 = \phi/n$ is always low the effective viscosity of an emulsion from $i+1$ droplets can be related to the effective viscosity from $i$ droplets via the Einstein formula $\eta_{i+1}/\eta_i = 1 + 2.5\gamma\phi_1$. Multiplying now these expressions yields for the effective viscosity of the emulsion $\overline{\eta}_{eff} = (1 + 2.5\gamma\phi/n)^n \to \exp(2.5\gamma\phi)$, where the last approximate expression is valid for large total number of droplets $n$. This non-linear dependence seems, however, less accurate than Eq. (5), since it does not diverge at $\gamma\phi \to 1$ as physically required.

### Transport coefficients and fluctuations

The role of fluctuations on transport coefficients is important for homogeneous systems as well as for composites. The problem can be reduced to a statistical analysis of the difference $\Delta\eta_{eff} \equiv \eta_{eff} - <\eta>$ as a function of $\Delta x \equiv x - <x>$, where $x$ indicates density, temperature or other thermodynamic parameters of the system. The average quantities, denoted by brackets, correspond to the equilibrium values. Considering the fluctuations as particles dispersed in a homogeneous medium, the logic of the previous section leads to Eq. (4) again

$$\gamma(\bar{\eta}_{eff}) = \gamma(\bar{\eta})\Phi \tag{6}$$

where $\Phi$ is the volume fraction of fluctuations and $\gamma$ is the interaction coefficient between the miscible regions with parameters $\eta$ and $<\eta>$.

In the frames of the linear non-equilibrium thermodynamics the steady state transport (thermal, diffusive, electrical, etc.) between a sphere and infinitive surrounding medium formulates one and the same boundary problem: determination of potential functions inside and outside the sphere with continuous fluxes across the boundary. To remind, the fluxes continuity takes place, when there is no physical boundary or if the real boundary shows negligible specific response. From the universality of the boundary problems it follows the universality of the relation between the interaction and transport coefficients [compare to Eq. (2)]

$$\gamma = \delta/(1+\delta) \tag{7}$$

where $\delta \equiv (\bar{\eta}-1)/a$ and $\bar{\eta} = \eta/<\eta>$. The factor $a$ is a number being 2.5 for viscosity [see Eq. (1)] and 3 for thermal conductivity, diffusion and dielectric permeability. Note that we do not consider here phase transitions and related surface tension effects.

In analogy with the previous section the system of Eqs. (6) and (7) give the following result of $\eta_{eff}$, developed in series of $\Delta\bar{\eta}$ up to the quadratic power term

$$\Delta\bar{\eta}_{eff} = \eta_{eff}/<\eta> - 1 = \Phi\Delta\bar{\eta} - \Phi(1-\Phi)(\Delta\bar{\eta})^2/a \tag{8}$$

On the other hand $\Delta\eta$ is related with the fluctuation parameter $\Delta x$ via the phenomenological dependence $\eta(x)$

$$\Delta\bar{\eta} = \eta/<\eta> - 1 = <\bar{\eta}'>\Delta x + <\bar{\eta}''>(\Delta x)^2/2 + \cdots \tag{9}$$

where the derivatives $<\bar{\eta}'>$ and $<\bar{\eta}''>$ refer to the equilibrium state. The combination of Eqs. (8) and (9) yields the following results for the first two average moments of $\Delta\bar{\eta}_{eff}$

$$<\Delta\bar{\eta}_{eff}> = \Phi[<\bar{\eta}''>/2 - (1-\Phi)<\bar{\eta}'>^2/a]<(\Delta x)^2> \tag{10}$$

$$<(\Delta\bar{\eta}_{eff})^2> = \Phi^2<\bar{\eta}'>^2<(\Delta x)^2> \tag{11}$$

At equilibrium $<\Delta x>=0$, while the dispersion $<(\Delta x)^2>$ is given by the corresponding thermodynamic relation.

We will apply the above results to two different systems to illustrate the role of fluctuations on the transfer in a composite and in a homogeneous system. Let us consider first viscosity fluctuations in an emulsion. Here the equilibrium corresponds to homogeneous droplet concentration with transport coefficient from Eq. (5) and $x \equiv \phi$. Expressing the derivatives in Eqs. (10) and (11) by Eq. (5) yields the following expressions for the leading terms at $a = 2.5$

$$<\Delta \eta_{eff}> = 2.5\Phi(1.5\gamma + \Phi/\phi)\gamma^2\phi <(\Delta\phi)^2> \qquad (12)$$

$$<(\Delta \eta_{eff})^2>^{1/2} = 2.5\Phi\gamma <(\Delta\phi)^2>^{1/2} \qquad (13)$$

The ratio $\Phi/\phi$ is kept in Eq. (12) as far its order of magnitude is unknown; note that the interaction coefficients at low Reynolds numbers are between 1 for solid and 2.5 for fluid spheres. The dispersion $<(\Delta\phi)^2>$ is related to the particles concentration $c$ and for dilute solutions one can express it as follows

$$<(\Delta\phi)^2> = v^2 <(\Delta c)^2> = \phi v/V \qquad (14)$$

where $v$ is the volume of a single droplet and $V$ is the average volume of a fluctuation region. As a second example let us describe the dielectric permittivity fluctuations in a homogeneous medium. The dependence of dielectric permittivity $\varepsilon$ as a function of the density $\rho$ in an equilibrium system is given by the Lorentz-Lorenz equation, which is analogical to Eq. (5),

$$\varepsilon = 1 + 3\alpha\rho/(1-\alpha\rho) \qquad (15)$$

where $\alpha$ is the molecular polarizability. As is known, the Lorentz-Lorenz equation is obtained also via the Maxwell approach and $\varepsilon$ is an effective quantity from microscopic point of view [11]. Following the procedures of the previous example, it is easy to obtain that

$$<\Delta \varepsilon_{eff}> = \Phi(\varepsilon+2)(\varepsilon-1)^2[2(\varepsilon-1)+\Phi(\varepsilon+2)]k_B T\kappa/27\varepsilon V \qquad (16)$$

Here the density fluctuations $<(\Delta\rho)^2> = k_B T \kappa <\rho>^2/V$ are presented via the isothermic compressibility $\kappa = (\partial \ln \rho/\partial p)_T$. In the both examples above the average fluctuating volume $V$ appears as a parameter, which will be discussed in the next section.

**Discussion**

The central assumption of the derivation of the effective viscosity in Eq. (5) is actually the dipole character at large distances of the perturbed field $\mathbf{v}'$ caused by a droplet placed in an external field $\mathbf{v}^0$ [see Eq. (1)]. A general question is about the range of validity of the relation between $\mathbf{v}'$ and $\mathbf{v}^0$ as a function of the droplet concentration. Of course, at higher particle concentration each droplet will be influenced by the field of the surrounding ones but in the case of a stationary homogeneous particles distribution the field can be only a deformation one according the Faxen theorem [7]. This is equivalent to an external field tensor $\mathbb{D}$ with the same characteristics (symmetry and zero trace) as $\mathbb{D}^0$. At a first look, the existence of this deformation field guarantees the validity of Eq. (5) at any particles concentration [note that the relation (5) is independent of $\mathbb{D}$], but the problem is that at higher concentrations the distances between particles are not large enough and the other faster decreasing components of $\mathbf{v}'$ cannot be neglected anymore. Thus the problem of the validity of Eq. (5) becomes equivalent with the validity range of dipole approximation (1).

The fluctuations as a source of non-homogeneities bring some specific problems in the transport theory, such as what is the order of the fraction $\Phi$ of fluctuations, the average volume of a fluctuation $V$, etc. We will restrict ourselves within the concept of complete correlation inside the volume $V$. The average concentration of fluctuations $C = \Phi/V$ appears in Eqs. (12) and (16). In the statistical physics the volume of a correlation domain reflects the interactions between the particles but the volume $V$ should also depend on the dissipative characteristics since the fluctuations are dissipative structures. For instance, in the case of independent particles (an ideal gas) the correlation length is of the order of the mean free path $\lambda \sim 1/\rho d^2$ [12], where $d$ is the diameter of the molecules and $\rho$ is the gas density. Due to isotropy the fluctuation volume reads $V \sim \lambda^3 \sim 1/\rho \phi^2$. Now we can estimate the dispersion from Eq. (14), $<(\Delta\phi)^2> \sim \phi^4$, and the first two moments of the effective viscosity from Eqs. (12) and (13), respectively, $<\Delta\bar\eta_{eff}> \sim \Phi\gamma^3\phi^5$ and $<(\Delta\bar\eta_{eff})^2> \sim \Phi^2\gamma^2\phi^4$. Hence, our analysis shows the presence of small but widely distributed viscosity fluctuations since $<(\Delta\eta_{eff})^2>/<\Delta\eta_{eff}>^2 \sim 1/\gamma^4\phi^6 > 1$. The only topic related to our problem is the effect of the Brownian motion of hard spheres ($\gamma = 1$) on the effective viscosity [13, 14]. The Batchelor result $\Delta\bar\eta_{eff} \sim \phi^2$ corresponds well to our expression for the viscosity root mean square variation $<(\Delta\bar\eta_{eff})^2>^{1/2}$, which is, however, expected since Batchelor considered only the linear dependence of the viscosity on the particle volume fraction. Note that $<\Delta\bar\eta_{eff}> = 0$ in this case.